\documentclass[aps,preprint,nofootinbib,showpacs]{revtex4}
\usepackage{graphicx,color,amsmath}
\newcommand{\be}{\begin{equation}}
\newcommand{\ee}{\end{equation}}
\newcommand{\ba}{\begin{eqnarray}}
\newcommand{\beq}{\begin{equation}}
\newcommand{\eeq}{\end{equation}}
\newcommand{\ea}{\end{eqnarray}}
\def\bea{\begin{eqnarray}}
\def\eea{\end{eqnarray}}
\begin{document}

\preprint{%
\vbox{%
\hbox{hep-ph/0606114}
}}
\title{
Associated production of a light pseudoscalar Higgs boson with a chargino
pair in the NMSSM.
}

\author{Abdesslam Arhrib$^1$, Kingman Cheung$^2$, Tie-Jiun Hou$^2$, 
Kok-Wee Song$^2$}
\affiliation{$^1$Facult\'e des Sciences et Techniques B.P 416 Tangier,
Morocco,} 
\affiliation{$^2$
Department of Physics and NCTS, National Tsing Hua University,
Hsinchu, Taiwan, R.O.C.
}

\date{\today}
\begin{abstract}
In the next-to-minimal supersymmetric standard model (NMSSM), the
unique $\lambda S H_u H_d$ in the superpotential gives rise to a
coupling involving the lighter pseudoscalar Higgs boson and a pair of
charged or neutral Higgsinos, even in the limit of zero mixing between
the two pseudoscalar Higgs bosons.  We study the associated production
of a very light pseudoscalar Higgs boson with a pair of charginos.
The novel signature involves a pair of charged leptons from chargino
decays and a pair of photons from the pseudoscalar Higgs boson decay,
plus large missing energy at the LHC and ILC.   The signal may help us
to distinguish the NMSSM from MSSM, 
provided that the experiment can resolve the two photons from the 
decay of the pseudoscalar Higgs boson.
\end{abstract}
\maketitle

\section{Introduction.}
Supersymmetry (SUSY) is one of the best motivated theories beyond
the standard model (SM).
Not only does it provide a natural solution to 
the gauge hierarchy problem, but also gives 
a dynamical mechanism for electroweak symmetry breaking and
a natural candidate for dark matter.
The most recent lower bound on the Higgs boson mass has been
raised to 114.4 GeV \cite{lep2}.  This in fact puts some stress on the soft
SUSY parameters, known as the little hierarchy problem, on
the minimal supersymmetric standard model (MSSM).
  Since the Higgs boson
receives radiative corrections dominated by the top squark loop, the mass bound
requires the top squark mass to be heavier than 1 TeV.  From the
renormalization-group (RG) equation of $M^2_{H_u}$, the magnitude of $M^2_{H_u}
\sim M^2_{\tilde{t}} \agt (1000\; {\rm GeV})^2$.  Thus, the parameters in the 
Higgs potential are fine-tuned at a level of a few percent in order to obtain
a Higgs boson mass of ${\cal O}(100)$ GeV.

Such fine-tuning has motivated a number of solutions to relieve the problem.
One of these is to add additional singlet fields to the
minimal supersymmetric standard model (MSSM).  The minimal version of the
latter is realized by adding a singlet Higgs field
to the MSSM, and becomes the next-to-minimal supersymmetric standard model
(NMSSM).  It has been shown \cite{jack} that in some corners of the parameter
space, the Higgs boson can decay into a pair of very light pseudoscalars
such that the LEP2 limit can be evaded. 
It has also been demonstrated that 
the fine-tuning or the little hierarchy problems are relieved \cite{jack}.
The NMSSM is in fact well motivated as it 
provides an elegant solution to the $\mu$ problem
in SUSY.  The $\mu$ parameter in the term
$\mu H_u H_d$ of the superpotential of the MSSM naturally has its value at
either $M_{\rm Planck}$ or zero (due to a symmetry).  However, the
radiative electroweak symmetry breaking conditions require the $\mu$ parameter
to be of the same order as $m_Z$ for fine-tuning reasons. Such a conflict
is coined as the $\mu$ problem \cite{mu}.  In the NMSSM, the $\mu$ term 
is generated dynamically through the 
vacuum-expectation-value (VEV), $v_s$,
of the scalar component of the additional Higgs field $S$,
which is naturally of the order of the SUSY 
breaking scale.
Thus, an effective $\mu$ parameter of the order of the electroweak scale
is generated.
Explicitly, the superpotential of the NMSSM is given by
\begin{eqnarray}
W &=& \mathbf{h_u} \hat{Q} \, \hat{H}_u \,\hat{U}^c
- \mathbf{h_d} \hat{Q}  \, \hat{H}_d \, \hat{D}^c
- \mathbf{h_e} \hat{L}  \, \hat{H}_d \, \hat{E}^c
 \nonumber \\
&& +\lambda \hat{S} \, \hat{H}_u \, \hat{H}_d+ \frac{1}{3}\kappa \, \hat{S}^3.
\end{eqnarray}
It is well-known that that the superpotential has a discrete $Z_3$ symmetry,
which may induce the harmful domain-wall effect \cite{domain}.
One possible way out is to introduce some nonrenormalizable operators
at the Planck scale that break the $Z_3$ symmetry through the harmless
tadpoles that they generate~\cite{PT}.

Once the domain wall problem is solved, the NMSSM is phenomenologically
very interesting.
With the additional singlet Higgs field, there are one more CP-even and
one more CP-odd Higgs bosons, and one more neutralino other than those
in the MSSM.  The Higgs phenomenology is much richer
\cite{jack,jack1,nmdecay,miller}, and so does the 
neutralinos \cite{neutral,vernon,hooper}.
One particular feature of the NMSSM is the allowable light pseudoscalar
boson $A_1$, which is consistent with existing data.  Since this
$A_1$  mainly comes from the singlet Higgs field, it can escape all the
experimental constraints when the mixing angle with the MSSM Higgs fields
goes to zero.  It was pointed out in Ref. \cite{jack} that even in the
zero mixing limit, there is always a SM-like Higgs boson that decays into
a pair of $A_1$'s, which helps the Higgs boson to evade the LEP bound.  
The possibility to detect such light pseudoscalar Higgs bosons coming 
from the $H_1$ decay was studied using the two photon mode of the $A_1$,
but the two photons may be too collimated for realistic detection
\cite{greg}.
Another possibility to detect such an almost decoupling
case is to search for the four tau-leptons coming from 
$h^0 \to A_1 A_1 \to 4 \tau$ decay \cite{pierce}.  Nevertheless,
in the large $\tan\beta$ and large $\langle v_s \rangle$ limits, the 
mixing angle is extremely small and approaching zero, such that the decay
of $A_1$ into tau-leptons or heavy quarks is negligible.

In this work, we probe another novel signature in the zero-mixing
limit.  The unique term $\lambda S H_u H_d$ in the superpotential
gives rise to the coupling of $\lambda S \tilde{H_u} \tilde{H_d}$,
which includes the neutral and charged Higgsinos.  We study the
associated production of a light pseudoscalar Higgs boson with a
chargino pair in the zero-mixing limit at the LHC and ILC. 
Provided that the pseudoscalar is very
light and the mixing angle is less than $10^{-3}$, the dominant decay
mode of $A_1$ is a pair of photons.  Thus, the novel signature for the
production is a pair of charged leptons and a pair of photons plus
large missing energy.  Such a signal can distinguish NMSSM from the
MSSM. 
We will show the production rates at the LHC and the
ILC.  One critical issue in identifying the two-photon decay of the light 
pseudoscalar Higgs boson is whether the two photons can be resolved.  
We will demonstrate the distribution of the opening angle between the
two photons, from which one can tell to what extent the experiments can
resolve the two photons.  The CMS detector has a ``preshower'' in the ECAL
that has the strong capability to resolve the two photons of the
neutral pion decay as it is the most important background for the 
intermediate mass Higgs boson search.  We can make use of this preshower
in the ECAL to resolve the two photons of the $A_1$ decay.  We will then
show the cross section for the associated production after imposing the
preshower requirements.  Once we resolve the two photons in the
decay of the pseudoscalar Higgs boson, we can differentiate the
NMSSM from the MSSM.

The organization is as follows.  In the next section, we describe the
particular region of parameter space in which the light pseudoscalar Higgs 
boson decouples and decays into a pair of photons.  In Sec. III, we calculate
the decay branching ratios of the $A_1$.  We then calculate the
associated production of the light pseudoscalar Higgs boson with a
chargino pair in the zero-mixing limit at the LHC and ILC in Sec. IV. 
We also work out the distribution of the opening angle of the photon pair.
We conclude in Sec. V.

\section{Zero mixing limit.}
The Higgs sector of the NMSSM consists of the usual two Higgs doublets 
$H_u$ and $H_d$ and an extra Higgs singlet $S$.
The extra singlet field is allowed to couple
only to the Higgs doublets of the model, the supersymmetrization of which
is that the singlet field only couples to the Higgsino doublets.
Consequently, the couplings of the singlet $S$ to gauge bosons and fermions
will only be manifest via their mixing with the doublet Higgs fields. 
After the Higgs fields take on the VEV's and rotating away the Goldstone
modes, we are left with a pair charged Higgs bosons, 3 real scalar fields,
and 2 pseudoscalar fields.  In particular, the mass matrix for
 the two pseudoscalar Higgs bosons $P_1$ and $P_2$ is 
\begin{equation}
V_{\rm pseudo} = \frac{1}{2} \;( P_1 \;\; P_2)  {\cal M}^2_P \; 
           \left( \begin{array}{c}
                             P_1 \\
                             P_2 \end{array} \right )
\end{equation}
with
\begin{eqnarray}
{\cal M}^2_{P\, 11} &=& M^2_A \;, \nonumber \\
{\cal M}^2_{P \, 12} &=& {\cal M}^2_{P \, 21} = \frac{1}{2} \cot\beta_s\,
\left(M^2_A \sin 2\beta - 3 \lambda \kappa v_s^2 \right ) \;, \nonumber \\
{\cal M}^2_{P \, 22} &=& \frac{1}{4} \sin 2\beta \cot^2 \beta_s
\,\left( M^2_A \sin 2\beta + 3 \lambda \kappa v_s^2 \right ) \nonumber \\
&& - \frac{3}{\sqrt{2}} \kappa A_\kappa v_s \;, 
\label{pmass}
\end{eqnarray}
where 
\begin{equation}
M_A^2 = \frac{\lambda v_s}{\sin 2\beta}\left(
             \sqrt{2} A_\lambda + \kappa v_s  \right ) \;,
\label{mA}
\end{equation}
and $\tan \beta = v_u/v_d$ and $\tan \beta_s = v_s/v$ and $v^2=v_u^2+v_d^2$.
Here $P_1$ is the one in MSSM while $P_2$ comes from the singlet $S$ and 
from the effects of rotating away the Goldstone modes. 
The pseudoscalar fields are further rotated to the diagonal basis
($A_1$, $A_2$) through a mixing angle
\cite{miller}:
\begin{eqnarray}
\left( \begin{array}{c} A_2 \\ A_1 \end{array} \right)=
\left( \begin{array}{cc}
     \cos \theta_A & \sin \theta_A \\
     -\sin \theta_A & \cos \theta_A  \end{array} \right)
\left( \begin{array}{c} P_1 \\ P_2 \end{array} \right)
\end{eqnarray}
where the masses of $A_i$ are arranged such that $m_{A_1} < m_{A_2}$.
At tree-level the mixing angle is given by
\begin{equation}
\tan \theta_A = \frac{ {\cal M}^2_{P\,12} }{ {\cal M}^2_{P\,11} - m^2_{A_1}}
 = \frac{1}{2} \cot\beta_s \,
  \frac{M^2_A \sin 2 \beta - 3 \lambda \kappa v_s^2}
  { M^2_A - m^2_{A_1} }
\label{tanA}
\end{equation}
In the approximation of large $\tan\beta$ and large $M_A$,
which is normally valid in the usual MSSM, 
the tree-level CP-odd masses can be written as \cite{miller}
\begin{eqnarray}
m_{A_2}^2 &\approx& M_A^2 \, (1+\frac{1}{4} \cot^2 \beta_s \sin^2 2\beta),
\\
m_{A_1}^2 &\approx& -\frac{3}{\sqrt{2}} \kappa v_s A_{\kappa}\;.
\label{small-ma}
\end{eqnarray}
We are interested in the case that $A_1$ is very light.  From 
Eq. (\ref{small-ma}) it can be seen that $m_{A_1}$ can be very small
if either $\kappa$ or $A_\kappa$ is very small, which is made possible
by a Pecci-Quinn (PQ) symmetry: $\kappa \to 0$ and $A_\kappa \to 0$.
We can achieve a small mixing angle by the cancellation between the
two terms in the numerator of Eq.(\ref{tanA}), by setting
\begin{equation}
  M_A^2 \sin 2 \beta - 3 \lambda \kappa v_s^2 = \sqrt{2} \lambda v_s 
\left( A_\lambda - \sqrt{2} \kappa v_s \right ) \approx 0
\qquad 
 \Rightarrow A_\lambda \approx \sqrt{2} \kappa v_s \;.
\label{cond}
\end{equation}
Explicitly, here we give a sample point in the parameter space:
\[
  \lambda = 1,\; v_s = 212 \;{\rm GeV},\; \kappa = 10^{-3},\;
  A_\lambda = 0.3 \; {\rm GeV},\; A_\kappa = -1 \;{\rm GeV} ,\;
 \tan\beta = 10\;,
\]
then it can give
\[
  m_{A_1} = 0.67 \;{\rm GeV}, \; \mu = 150\;{\rm GeV}, \; 
 \tan \theta_A = 0.5 \times 10^{-4} \;.
\]

We have the following parameters in the NMSSM 
in addition to those of the MSSM:
$\lambda$, $\kappa$, $A_\lambda$, $A_\kappa$, and $v_s$
($m_S^2$ has been eliminated by one of the tadpole equations in the
electroweak symmetry breaking.)  Since $\lambda v_s/\sqrt{2} = \mu$, we 
can use $\mu$ and $\lambda$ in place of $v_s$.  Also from 
Eqs. (\ref{mA}), (\ref{small-ma}) and (\ref{tanA}) we can trade
$\kappa$, $A_\lambda$, and $A_\kappa$ for $m^2_A$, $m_{A_1}$ and 
$\sin\theta_A$.  
The small $m_{A_1}$ can be achieved by requiring $\kappa \to 0$ in
Eq. (\ref{small-ma}) 
while keeping $v_s$ and $A_\kappa$ typical.  The small mixing
angle, on the other hand, is achieved by the 
condition in Eq. (\ref{cond}) such
that the fine-tuned cancellation is possible to give a small value of
$\tan\theta_A$.  We admit that this is a fine tuning requirement.
Recall in the CP violating supersymmetry a natural mechanism to
suppress the contributions from various sources of CP violating phases
is to have a cancellation among various sources \cite{nath}.  If the
measurement of EDM continues to push to more stringent limits, 
more and more fine-tuned cancellations are 
needed to suppress the SUSY contributions.

\begin{figure*}[th!]
\centering
\includegraphics[width=3.5in,clip]{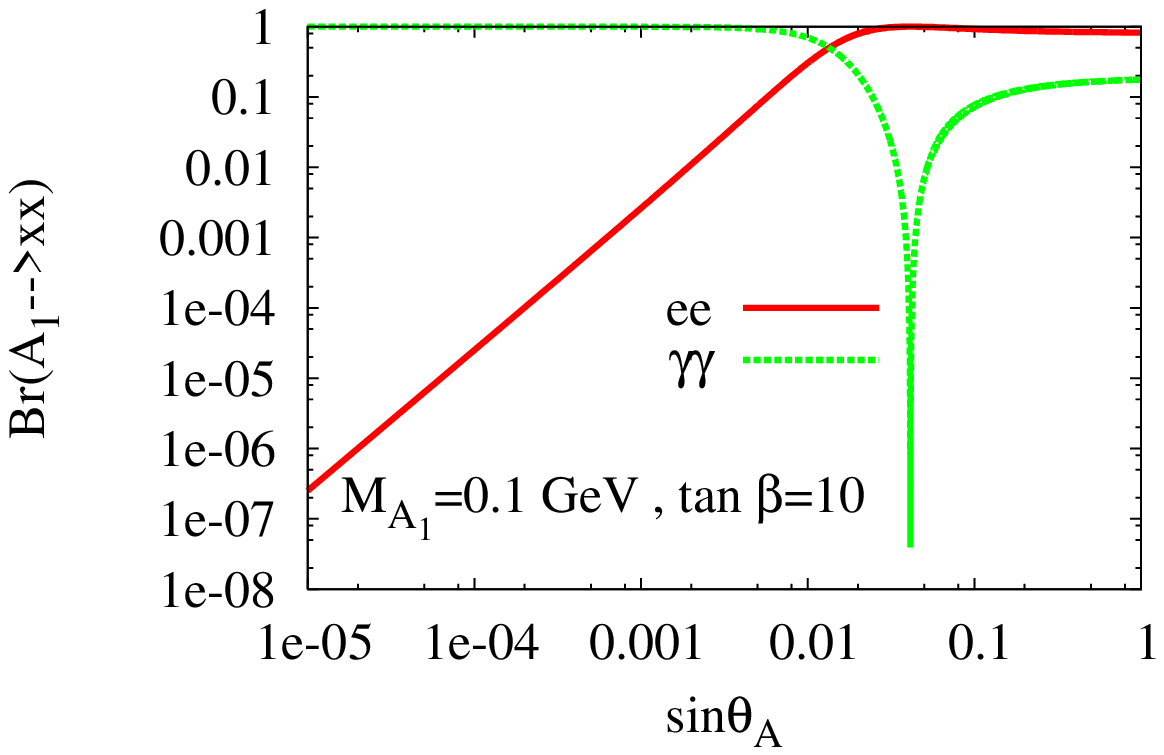}
\includegraphics[width=3.5in,clip]{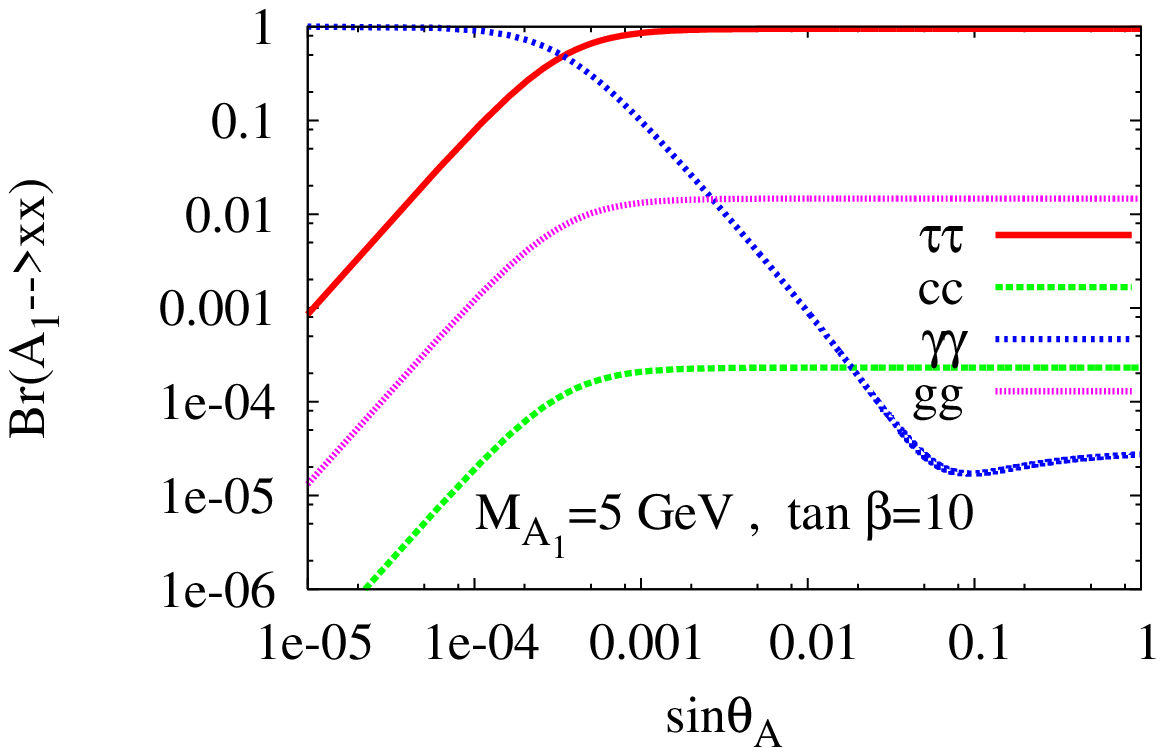}
\caption{\label{fig1} \small
Decay branching ratios for the light pseudoscalar Higgs boson
versus the mixing angle $\sin\theta_A$ for $\lambda=1$, $\mu=150$, $M_2=500$
GeV. (a) $m_{A_1} = 0.1$ GeV and (b) $m_{A_1}=5$ GeV.}
\end{figure*}

\section{Decay.}
A lot of existing constraints on the lightest pseudoscalar Higgs boson
$A_1$ depend on the mixing angle $\sin\theta_A$.  When $\sin\theta_A$ goes
to zero, the $A_1$ decouples and behaves like the singlet.  This light
$A_1$ can be extremely light without violating any existing data.  It was
pointed out \cite{jack} that it can be produced in the scalar Higgs boson
decay, $H_1 \to A_1 A_1$, which is due to the term
$\lambda \hat S \hat H_u \hat H_d$ in the superpotential.
We found in this work that there is another novel signature 
for this light $A_1$ from the same term.
We calculate
the associated production of $A_1$ with a pair of charginos, followed
by $A_1 \to \gamma\gamma$ decay at hadronic and $e^+ e^-$ colliders.
This is an undebatable signal of the decoupling regime of the NMSSM.

In the limit of zero mixing, the $A_1$ only couples to a pair of charginos
and neutralinos.  Therefore, the dominant decay mode is
$\gamma\gamma$ via a chargino loop if $m_{A_1}$ is very light.
When we turn on the small mixing angle,
other modes, such as $q\bar q$, $\ell^+ \ell^-$, and $gg$, appear, which
will eventually dominate when the mixing angle is larger than $O(10^{-3})$.
We show a typical decay branching ratio versus the mixing angle for 
$m_{A_1} = 0.1, 5$ GeV in Fig. \ref{fig1}.
When $m_{A_1}$ is as light as 0.1 GeV, only the $e^+ e^-$ and $\gamma\gamma$
modes are possible. The $e^+ e^-$ mode scales as $\sin^2\theta_A$, and so
the $e^+ e^-$ mode increases sharply as $\sin\theta_A$ increases in 
Fig. \ref{fig1}(a).  
As $m_{A_1}$ increases other $f\bar f$ modes open up, such
as $\tau^+\tau^-, c\bar c, gg$.  As long as $\sin\theta_A \alt 10^{-3}$
the $\gamma\gamma$ dominates the decay of $A_1$.  

\section{Associated Production.}
The coupling of $A_i$ to  charginos comes from the usual Higgs-Higgsino-gaugino
source and, specific to NMSSM, from the term $\lambda \hat S \hat H_u \hat
H_d$ in the
superpotential.  The interaction is given by
\begin{eqnarray}
{\cal L}_{A \chi^+ \chi^+ } &=&
i \overline{\widetilde{\chi}^+_i} \left ( C_{ij} P_L - C_{ji}^* P_R \right)
            \widetilde{\chi}^+_j\, A_2
 \nonumber \\ 
&+&
i \overline{\widetilde{\chi}^+_i} \left ( D_{ij} P_L - D_{ji}^* P_R \right)
            \widetilde{\chi}^+_j\, A_1 \;,
\label{aichij}
\end{eqnarray}
where
\begin{eqnarray}
C_{ij} &=& \frac{g}{\sqrt{2}} \left( \cos\beta \,\cos\theta_A \,U^*_{i1} \,
V^*_{j2} + \sin\beta \, \cos\theta_A \, V^*_{j1} \, U^*_{i2} \right )
   \nonumber \\
&& - \frac{\lambda}{\sqrt{2}} \sin\theta_A \, U^*_{i2} V^*_{j2}
 \;, \nonumber \\
D_{ij} &=& \frac{g}{\sqrt{2}} \left( -\cos\beta \,\sin\theta_A \,U^*_{i1} \,
V^*_{j2} - \sin\beta \, \sin\theta_A \, V^*_{j1} \, U^*_{i2} \right )
  \nonumber \\
&& - \frac{\lambda}{\sqrt{2}} \cos\theta_A \, U^*_{i2} V^*_{j2}
 \;,
\end{eqnarray}
where $P_{L,R}=(1 \mp \gamma_5)/2$ are the chiral projectors.

\begin{figure}[t!]
\centering
\includegraphics[width=3.5in]{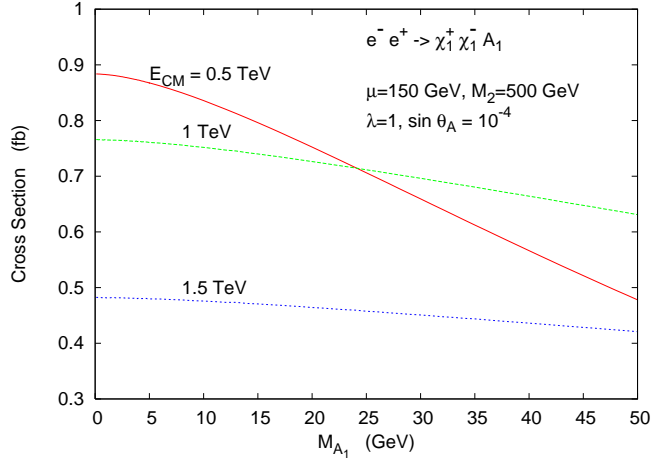}
\caption{\label{fig2} \small
Production cross sections for $e^- e^+ \to \widetilde{\chi}^+_1
\widetilde{\chi}^-_1 A_1$ at the ILC with $\sqrt{s}=0.5,\, 1,\, 1.5$ TeV.
We have chosen $\lambda=1$, $\sin\theta_A =10^{-4}$, $\mu = 150$ GeV, 
$M_2 = 500$ GeV, and $\tan\beta = 10$.}
\end{figure}

We stress in passing that in the limit of zero mixing, 
the production of $A_1$ 
through the Drell-Yan process $e^+e^-/pp\to A_1 \Phi$ is very suppressed.
So is the gluon fusion since only quarks 
can mediate inside the loops.
The associated production of $A_1$ with a chargino pair proceeds via the
Feynman diagrams, in which the $A_1$ radiates off the chargino legs. The 
radiation off the intermediate $Z$ is not considered in the very small
mixing limit.  Note that the production is proportional to
$|\lambda \cos\theta_A U_{12} V_{12} |^2$, which implies a large Higgsino
component in $\widetilde{\chi}_1^+$ is necessary for large cross sections.
Details of the calculation will be given in a future publication.
We choose $\mu = 150$ GeV and a much larger $M_2 =500$ GeV, 
$\lambda = 1$
\footnote{
We have considered the largest possible value of $\lambda$, which
should be of the order of $O(1)$.  The size is limited by the
perturbativity argument when the Yukawa coupling is evolved to the GUT
scale.  We have followed the prescription from other papers.  In
general, the perturbativity argument is rather loose.  There may be
some other new physics that appear well below the GUT scale.
But roughly $\lambda \approx O(1)$ is the upper limit applied to 
$\lambda$ for the perturbativity reason.
}
 and $\sin\theta_A = 10^{-4}$ in our results.  
We show the production cross sections  at $e^+ e^-$ colliders 
versus $m_{A_1}$ for $\sqrt{s}=0.5,1,1.5$ TeV in Fig. \ref{fig2}.
Note that the cross section is insensitive to $\sin\theta_A$ as long as
it is less than $10^{-2}$. Also, in this near-zero mixing region, the
cross section scales as $\lambda^2$.  With $O(500)$ fb$^{-1}$ yearly
luminosity at the ILC, the number of raw events is of the order of
$O(500)$.
The signature is very spectacular with a pair of charged leptons and 
a pair of photons with a large missing energy.  In contrast to the process
of $h\to A_1 A_1 \to 4 \gamma$ \cite{greg}, 
the photon pair is less collimated because
the $A_1$ radiating off the chargino would not be as energetic as the
$A_1$ from Higgs decay.  Almost all SM backgrounds are reducible once the
photon pair and the charged lepton pair are identified together with 
large missing energies.

The leptonic branching ratio of the chargino can increase if the slepton
or sneutrino mass is relatively light.  
One can also increase the detection rates by including the hadronic decay 
of the charginos.  Therefore, in the final state we can have 
(i) two charged leptons $+$ two photons $+$ $\not\!{E}_T$, 
(ii) one charged leptons $+$ two jets $+$ two photons $+$ $\not\!{E}_T$, or
(iii) 4 jets $+$ two photons $+$ $\not\!{E}_T$.

\begin{figure}[t!]
\centering
\includegraphics[width=3.4in]{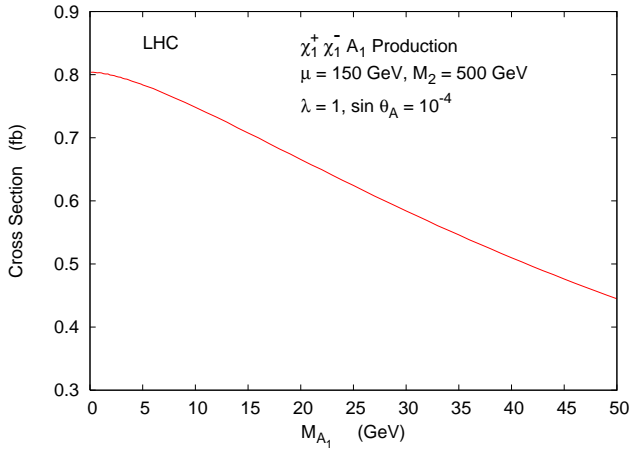}
\caption{\small \label{fig3}
Production cross sections for $pp  \to \widetilde{\chi}^+_1
\widetilde{\chi}^-_1 A_1$ at the LHC.
We have chosen $\lambda=1$, $\sin\theta_A =10^{-4}$, $\mu = 150$ GeV, 
$M_2 = 500$ GeV, and $\tan\beta = 10$.
}
\end{figure}

We show the production
cross section for $pp \to \widetilde{\chi}^+_1
\widetilde{\chi}^-_1 A_1$ at the LHC in Fig. \ref{fig3}, 
with the same set of parameters as in Fig. \ref{fig2}.  We obtain a cross
section slightly shy of $O(1)$ fb.  With a yearly luminosity of 100 fb$^{-1}$
one can have about $O(100)$ raw events.  It remains possible to detect the
pseudoscalar Higgs boson and chargino decays.  Experiments
can search for the same final states that we listed for the ILC.  Almost 
all SM backgrounds are reducible if the photon pair can be resolved
and measured,  together with the charged leptons or jets plus 
missing energies.

\begin{figure}[t!]
\includegraphics[width=5in]{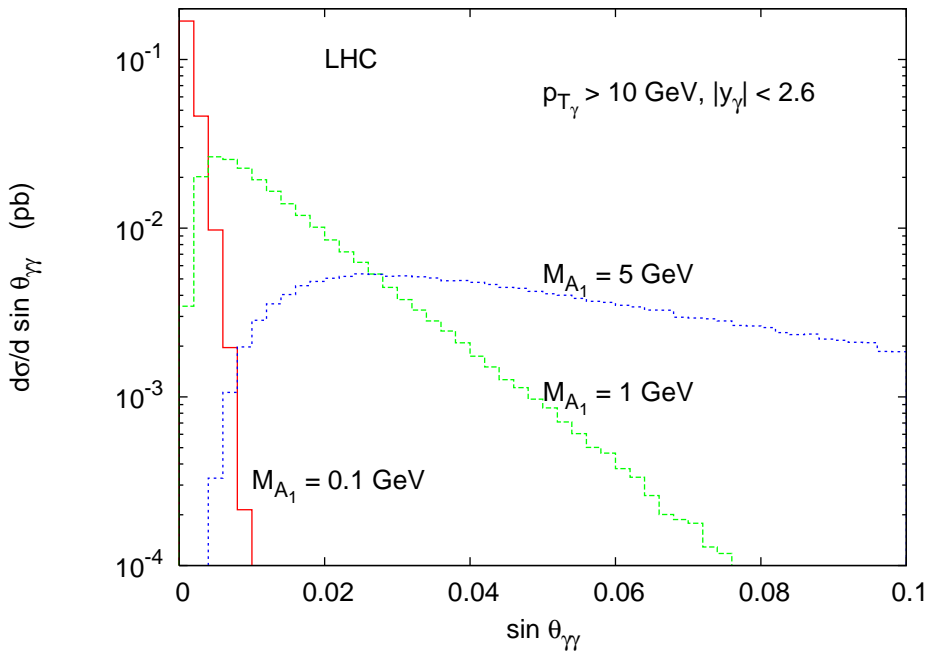}
\caption{\small \label{angle}
The differential cross section versus the sine of the opening angle
between the two photons for $\lambda = 1$ and $\sin\theta_A = 10^{-4}$ 
at the LHC.  Requirements of $p_{T\gamma} > 10$ GeV and $|y_\gamma| < 2.6$
are imposed.
}
\end{figure}

The critical issue here is whether the LHC experiment
 can resolve the two photons 
in the decay of the 
pseudoscalar Higgs boson.  We perform a monte carlo study for the 
production of $\widetilde{\chi}^+_1 \, \widetilde{\chi}^-_1\, A_1$ 
followed by the decay of 
$A_1 \to \gamma \gamma$.  Since $A_1$ is a pseudoscalar, it is sufficient
to study the $2\to 2$ phase-space decay of $A_1$.  We impose transverse
momentum and rapidity requirements on the photons:
\begin{equation}
p_{T_\gamma} > 10 \; {\rm GeV}\,, \qquad |y_\gamma| < 2.6 \,,
\end{equation}
which are in accord with the ECAL of the CMS detector \cite{cms}.  The 
resolution of the ``preshower'' detector
quoted in the report is as good as $6.9$ mrad.  We shall use
10 mrad as our minimum separation of the two photons that the detector
can resolve.
We show the distribution of the sine of the opening
 angle between the two photons
for $M_{A_1} = 0.1,\; 1,\; 5$ GeV in Fig. \ref{angle}.  It is easy to
understand that for  $A_1$ as light as $0.1$ GeV all the cross sections are
within the opening angle $\theta_{\gamma\gamma} < 0.01$ rad.  When $M_{A_1}$ 
increases to 1 GeV, more than half of the cross sections are beyond 0.01 rad.
For $M_{A_1}$ as large as 5 GeV almost all cross sections are beyond 
$\theta_{\gamma\gamma} > 0.01$ rad.  We show the resultant cross sections
for $M_{A_1} = 0.1 - 5 $ GeV 
with $p_{T_\gamma} > 10 \; {\rm GeV}$, $|y_\gamma| < 2.6$, and 
$\theta_{\gamma\gamma}> 0.01$ rad in Table \ref{table}.  Suppose the LHC
can accumulate $O(500 - 1000)$ fb$^{-1}$ luminosity, so $M_{A_1}$ as low as 
$0.3 - 0.4$ GeV are possible to be detected.
For a mere $O(100)$ fb$^{-1}$ luminosity, the size of the cross section
in Table I shows that it is only possible to detect $m_{A_1} > 1 $ GeV.

The final issue is the background suppression.  We have shown in
Fig. \ref{angle} that for $m_{A_1} \sim 0.1$ GeV, almost all cross
 section lies
below $\theta_{\gamma\gamma} < 0.01$, which is our conservative choice
of resolution according to the preshower detector of the CMS.
However, when $m_{A_1} \agt 1$ GeV, more than half of the cross
section survives this $\theta_{\gamma\gamma} > 0.01$ cut.  We can also
reconstruct the invariant mass of the photon pair to identify the
pseudoscalar Higgs boson $A_1$ and separate it from the other SM mesons
such as $\pi^0$ and $\eta$.  Photon and lepton isolation cuts are the
most useful ones to reject the jet-faking background and other QCD
background.  The remaining backgrounds are mostly gauge-boson pair and
$t \bar t$ plus photons/jets production with the photons/jets
radiating off fermion or gauge boson legs.  Although they are
irreducible, they are of higher order in couplings and should be small.
Perhaps, the more serious background issue in the LHC environment may
be the combinatorial background because of many photons within a jet.
Again, using strong photon-pair isolation (that is without hadronic
jets around the photon pair) one should be able to substantially
reduce this background.


\begin{table}[b!]
\caption{\label{table} \small
Cross sections in fb for associated production of 
$\widetilde{\chi}^+_1 \; \widetilde{\chi}^-_1 \; A_1$ followed
by $A_1 \to \gamma\gamma$.  The cuts applied to the two photons are:
$p_{T_\gamma} > 10$ GeV, $|y_\gamma| < 2.6$, and $\theta_{\gamma\gamma} > 
10$ mrad.  }
\medskip
\begin{ruledtabular}
\begin{tabular}{cc}
$M_{A_1}$  ( GeV)  &  Cross Section  (fb) \\
\hline
$0.1$                &  $0.0$ \\
$0.2$                &  $0.011$ \\
$0.3$                &  $0.0405$  \\
$0.4$                &  $0.078$   \\
$0.5$                &  $0.12$  \\
$1$                &  $0.26$  \\
$2$                &  $0.38$  \\
$3$                &  $0.42$  \\
$4$                &  $0.44$  \\
$5$                &  $0.44$  
\end{tabular}
\end{ruledtabular}
\end{table}

\section{Discussions and Conclusions.}
One may ask if a very light pseudoscalar Higgs boson is consistent
with the muon anomalous magnetic moment ($g-2$) because it can contribute
substantially to $g-2$ at both 1-loop and 2-loop levels.  However, it
was shown that the 2-loop Barr-Zee type contributions with a
light pseudoscalar can be of comparable size as the 1-loop
contributions and opposite in sign \cite{ck}.  
Note that the contributions of the light $A_1$ of the NMSSM go
to zero as $\sin\theta_A \to 0$.
In the NMSSM, there could also be a light neutralino \cite{hooper}
that can contribute to $g-2$.
In addition, 
there are many parameters in the MSSM, such as gaugino and sfermion 
masses, which the $g-2$ depends on.
Thus, one can carefully take into account both 1- and 2-loop contributions
and by adjusting the NMSSM parameters, such that 
the $g-2$ constraint is satisfied.
There are other constraints on a light pseudoscalar from rare $K$ and
$B$ meson decays, such as $b\to s A_1$ and $s\to d A_1$,
$B-\overline{B}$ mixing, $B_s \to \mu^+ \mu^-$, and $\Upsilon \to A_1
\gamma$ \cite{hiller,hooper}.  However, it is obvious that in these
processes the light pseudoscalar interacts via the mixing with
the MSSM pseudoscalar.  Thus, in the limit of zero-mixing the constraints
on the light $A_1$ can be easily evaded.

The major difference between MSSM and NMSSM is the existence of a singlet
field, which gives rise to a scalar, a pseudoscalar, and a neutralino, 
in addition to the particle contents of the MSSM.  We have shown that
it is possible to have a very light pseudoscalar with a tiny mixing
with the MSSM pseudoscalar.  Such a light pseudoscalar boson is consistent 
with all existing constraints.  
The discovery mode has been shown \cite{jack} to be $H \to A_1 A_1$, which
enjoys a large production cross section.  However, the photon pair from
the $A_1$ decay may be too collimated.
In this paper, we have pointed out another unambiguous 
signature from the associated production of the light pseudoscalar with
a pair of charginos at the LHC and ILC, with 
a pair of charged leptons and a pair of photons plus large missing energy
in the final state.  
We have also shown that the event rates at the LHC and ILC should be 
enough to identify such a signature when $M_{A_1}$ is larger than 
$1$ GeV.


\begin{acknowledgments}
The research was supported by the NSC of Taiwan under
Grant 94-2112-M-007-010-. We thank Wai-Yee Keung, We-Fu Chang, and 
Francesca Borzumati for discussion.
\end{acknowledgments}


\end{document}